\documentclass[9pt,twocolumn]{article}

\usepackage[utf8]{inputenc}
\usepackage[T1]{fontenc}
\usepackage{times}
\usepackage[a4paper,margin=1in]{geometry}

\usepackage{booktabs}
\usepackage{cite}
\usepackage{amsmath,amssymb,amsfonts}
\usepackage{algorithmic}
\usepackage{graphicx}
\usepackage{subcaption}
\usepackage{textcomp}
\usepackage{xcolor}
\usepackage{siunitx}
\usepackage{multirow}
\usepackage{hyperref}

\graphicspath{{figures/}}

\definecolor{darkgreen}{RGB}{0,120,0}

\newcommand{\todo}[1]{}
\newcommand{\review}[1]{}

\ifdefined\autocite
\else
  \newcommand{\autocite}{\cite}
\fi

\def\BibTeX{{\rm B\kern-.05em{\sc i\kern-.025em b}\kern-.08em
    T\kern-.1667em\lower.7ex\hbox{E}\kern-.125emX}}

\begin{document}

\title{Validation of a Software-Defined 100-Gb/s RDMA Streaming Architecture for Ultrafast Optoacoustic and Ultrasound Imaging}

\author{
Federico Villani$^{1}$, Christian Vogt$^{2}$, Luca Specht$^{1}$, Jero Schmid$^{1}$, Xiang Liu$^{3,4}$,\\ 
Andrea Cossettini$^{1}$, Daniel Razansky$^{3,4}$, Luca Benini$^{1,5}$\\[1ex]
\small $^{1}$Integrated Systems Laboratory (IIS), ETH Z{\"u}rich, Switzerland\\
\small $^{2}$Center for Project Based Learning (PBL), ETH Z{\"u}rich, Switzerland\\
\small $^{3}$Institute for Biomedical Engineering, ETH Z{\"u}rich, Switzerland\\
\small $^{4}$Institute of Pharmacology and Toxicology and Institute for Biomedical Engineering, University of Z{\"u}rich, Switzerland\\
\small $^{5}$Department of Electrical, Electronic, and Information Engineering (DEI), University of Bologna, Italy\\[1ex]
\small Contact: villanif@iis.ee.ethz.ch
}

\date{}

\maketitle

\begin{abstract}
Optoacoustic (OA) imaging has emerged as a powerful investigation tool, with demonstrated applicability in oncology, neuroscience, and cardiovascular biology. However, its clinical translation is limited with the existing OA systems, which often rely on bulky and expensive acquisition hardware mainly optimized for pulse-echo ultrasound (US) imaging. Despite the fact that OA imaging has different requirements for receive bandwidths and timing synchronization with external laser sources, there is a strong need for unified OA-US imaging platforms, as pulse-echo US remains the standard tool for visualizing soft tissues. 
To address these challenges, we propose a new data acquisition architecture for ultrafast OA and US imaging that fully covers the requirements for large channel counts, wide bandwidth, and software-defined operation. LtL combines state-of-the-art wideband analog front-ends, a Zynq UltraScale+ MPSoC integrating FPGA fabric with an Application Processing Unit, and a 100 GbE Remote Direct Memory Access (RDMA) backend enabling raw-data streaming at up to 95.6 Gb/s. The architecture avoids local buffers followed by burst transfers, which commonly constrain sustainable frame rate and recording intervals, thus achieving true continuous and sustained streaming of raw data. We validate the core elements of the LtL architecture using a 16-channel demonstration system built from commercial evaluation boards. We further verify the signal chain for up to 256-channel scalability, confirming the wide bandwidth capabilities to support state-of-the-art data transmission speeds.
\end{abstract}


\section{Introduction}
\label{sec:introduction}

Optoacoustic (OA) imaging has emerged as a powerful technique using pulsed optical excitation and broadband acoustic detection to generate images based on spatial variations in optical absorption \cite{manoharPhotoacousticsHistoricalReview2016}. Modern OA tomography has demonstrated significant potential in biomedical research, with applications spanning oncology \cite{mcnallyCurrentEmergingClinical2016}, neuroscience \cite{naPhotoacousticComputedTomography2021}, and cardiovascular biology \cite{liuFunctionalPhotoacousticMicroscopy2022}. 
However, the clinical translation of OA technology still lags behind its true potential \cite{parkClinicalTranslationPhotoacoustic2025,wenClinicalPhotoacousticUltrasound2022}, with the major barriers being the high cost and bulky form factor of the optical excitation and data acquisition (DAQ) subsystems. In fact, high-performance OA tomography systems require DAQ front-ends capable of digitizing a large number of broadband, low-amplitude receive channels for every emitted laser pulse. To enable signal readout, many OA systems rely on conventional US platforms \cite{pangMultimodalityPhotoacousticUltrasound2021}, which are optimized for pulse-echo operation and struggle to meet OA-specific requirements such as ultrawide receive bandwidth, high input impedance, and precise synchronization to external laser sources.

A growing need further exists for developing multi-modal imaging systems integrating OA imaging with pulse-echo ultrasound (US), as the latter remains the standard for clinical diagnostics of soft tissues and can provide excellent anatomical reference for functional OA data~\cite{bouchardUltrasoundguidedPhotoacousticImaging2014}. In this context, US itself is undergoing rapid innovation, with ultrafast US techniques enabling kHz frame rates for functional and quantitative imaging. Yet, most ultrasound platforms remain proprietary, not fully programmable, and heavily rely on fixed-function circuitry. Openness and flexibility at both software and hardware levels appear as critical requirements to rapidly prototype and evaluate novel acquisition and reconstruction schemes \cite{jonveauxReviewCurrentSimple2022,boniUltrasoundOpenPlatforms2018}.

Another challenge shared by OA and US research systems is the lack of true continuous streaming of raw data. Only recently a small number of research and commercial platforms have enabled high-throughput streaming interfaces alongside their internal beamforming engines \cite{lagonigroModularScalableSystem2025, verasonicsVantageNXTResearch, opensonicsUltrasoundResearchProducts}, with most systems still relying on local buffering in onboard memory, followed by burst transfers to a host. This approach constrains both the maximum sustainable frame rate and the maximum record length, severely limiting the study of fast biodynamic processes evolving over longer timescales where uninterrupted image sequences at high sampling rates are required.

Taken together, these requirements define a design space that cannot be matched with existing hardware designs. To the best of our knowledge, no cost-effective, open, and integrated OA-US imaging platform exists that can support high channel counts and wideband acquisition at ultrafast frame rates over long timespans without being limited by buffer sizes. Such a platform should also remain flexible enough to be programmed entirely in software without requiring expertise in hardware description languages.

To address these challenges, we propose \textit{ListenToLight~(LtL)}, a software-defined 256-channel architecture for ultrafast OA and US imaging. LtL combines state-of-the-art (SoA), wideband analog front-ends (for up to 125 MSPS sampling rates), a Zynq Ultrascale+ MPSoC that tightly integrates FPGA fabric with an Application Processing Unit (APU) to enable software-based triggering, sequencing, and system configuration, and a 100 GbE Remote Direct Memory Access (RDMA) backend enabling sustained raw-data streaming at up to 95.6 Gbps.

Here we validate the core elements of the LtL architecture using a 16-channel demonstration system built from commercial evaluation boards. Although only sixteen analog channels are present, the digital backend operates at 256-channel datapath width, enabling us to evaluate the complete data pipeline under realistic throughput conditions of a high-end OA tomography system. Ultrasound and optoacoustic phantom experiments confirm precise timing, correct data handling, and the practical feasibility of the proposed architecture.

The paper is organized as follows: Sect.~\ref{sect:related} reviews related works in OA and US research platforms; Sect.~\ref{sect:system_concept} presents the LtL system concept, including data, clocking interfaces, and control interfaces; Sect.~\ref{sect:sixteen} describes the 16-channel demonstration platform; Sect.~\ref{sect:characterization} and Sect.~\ref{sect:validation_phantom} report hardware characterization and experimental validation; Sect.~\ref{sect:soa} compares LtL with related state-of-the-art works; and Sect.~\ref{sect:conclusion} discusses scalability and future directions.

\section{Related Works}
\label{sect:related}

\subsection{Hybrid Optoacoustic-Ultrasound Systems}

Several preclinical and clinical systems have demonstrated the potential of combining OA and US imaging.
Early OA--US systems typically relied on commercial US scanners for the backend signal acquisition, with external laser triggers integrated in the timing chain. Although these platforms provided mature beamforming, they were optimized for fixed pulse--echo workflows and lacked the flexibility required for broadband high-quality OA data acquisition. In particular, relevant examples were typically limited to 128-element linear array systems without access to raw data or low-level timing control, which rendered poor quality OA images due to severe limited-view and out-of-plane artifacts~\cite{nguyenThreedimensionalViewOutofplane2020}. The closed firmware architectures and tight coupling to vendor-specific hardware further hindered the integration of high-repetition-rate lasers, making these systems unsuitable for real-time dual-mode imaging \cite{xuPhotoacousticUltrasoundDualmodality2012,daoudiHandheldProbeIntegrating2014}.
    
Fully custom hybrid OA–US designs have addressed some of these limitations by integrating both modalities on dedicated hardware.
Mer{\v{c}}ep et al.~\cite{mercepWholebodyLiveMouse2015} presented a preclinical hybrid imaging scanner for whole-body preclinical studies, further incorporating rapid laser wavelength tuning for real-time spectroscopic data acquisition.
Robin et al.~\cite{robinDualModeVolumetricOptoacoustic2022} reported dual-mode volumetric imaging and contrast-enhanced US by employing a custom data acquisition platform (Falkenstein Mikrosysteme GmbH, Germany). Other works developed custom FPGA-based backends for intravascular OA and US imaging, focusing on compact implementations \cite{wuFPGABasedBackendSystem2019}.

\begin{figure}
    \centering
    \includegraphics[width=0.9\columnwidth]{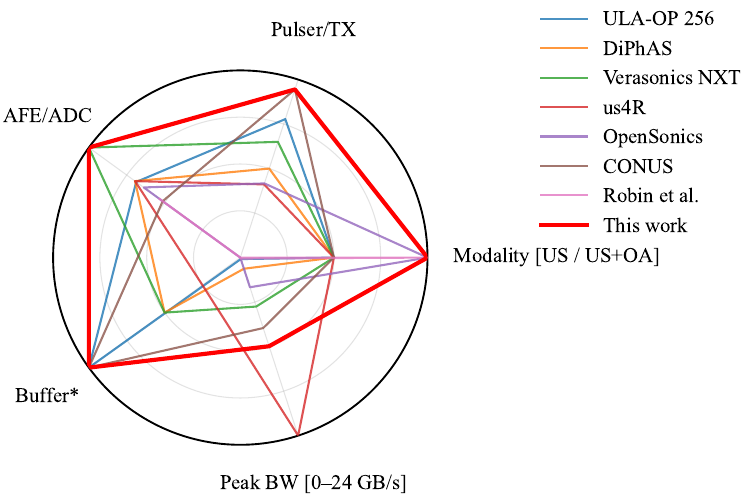}
    \caption{Radar plot comparing key performance metrics of LtL against state-of-the-art (SoA) hybrid OA--US imaging platforms (see Table~\ref{tab:platform-comparison}). Axes are normalized to $[0,1]$. Modality: $US=0.5 OA+US=1$ . Pulser: $s_{\mathrm{TX}}=\tfrac{1}{3}s_{\mathrm{levels}}+\tfrac{1}{3}s_{V_{\mathrm{pp}}}+\tfrac{1}{3}s_{f_{\max}}$. AFE/ADC: $s_{\mathrm{AFE}}=\tfrac{1}{2}s_{N_{\mathrm{bits}}}+\tfrac{1}{2}s_{f_s}$. Buffer*: max on-system memory; for host-buffered systems: equivalent to max. on-system buffer.}
    \label{fig:soa_radar}
\end{figure}

While prior works demonstrated successful OA-US integration, major limitations remained in terms of limited scalability, timing and system control fixed at hardware level, and custom bulky hardware designs with limited openness. In addition, no fast data interfaces are available that can support kHz frame rate imaging with large channel counts, as fast acquisitions are typically buffered into local memory with only short-duration transmission bursts. LtL addresses all these limitations as shown in the radar plot in Fig.~\ref{fig:soa_radar}: we provide a quantitative comparison to SoA and a detailed description of the chosen metrics in the later Sect.~\ref{sect:soa}.

\subsection{Open US research platforms}

In parallel to hybrid OA-US systems, multiple open US research platforms have emerged \cite{boniUltrasoundOpenPlatforms2018}. In the following, we summarize key systems for research use, i.e., offering access to raw data and varying degrees of low-level programmability.

\paragraph*{Open research scanners with high-end embedded processing}

Open platforms typically combine multiple transmit/receive (TX/RX) channels with beamforming and real-time processing in field-programmable gate arrays (FPGAs) or digital signal processors (DSPs). Additionally, they typically stream beamformed or raw data to a host computer for storage and post-processing~\cite{boniUltrasoundOpenPlatforms2018}.

ULA-OP \cite{tortoliULAOPAdvancedOpen2009} and its successor ULA-OP 256 \cite{boniULAOP256256Channel2016} are representative examples. ULA-OP 256 provides 256 fully-programmable channels with modular front-ends, on-board FPGAs and DSPs for real-time high-frame-rate imaging, as well as open access to firmware and processing pipelines for implementing new imaging methods.
A more recent advancement by the same research group is CONUS \cite{lagonigroModularScalableSystem2025}, a scalable, modular system built from synchronized 64-channel front-ends, each including integrated pulsers, 50 MHz analog front-ends (AFEs), and FPGAs (10CX220F780, Altera). In CONUS, digitized data are streamed via 40 Gigabit Ethernet optical links to GPU-based back-end systems, reaching 25Gbit/s out of a single module and a peak of 76 Gbit/s of cumulative bandwidth \cite{vignoliHighspeedDataTransfer2025}. These platforms demonstrate that modular front-ends and open firmware can support advanced imaging modes and large effective channel counts, but are still limited in the sampling frequencies (hindering multiple high-frequency OA applications \cite{schwarzImplicationsUltrasoundFrequency2015}) and are bandwidth-limited by the network interface link.

SARUS is another example of a high-end synthetic-aperture research platform that offers high channel counts, flexible transmit sequencing, and real-time processing implemented in dedicated hardware~\cite{jensenSARUSSyntheticAperture2013}. The DiPhAS system, developed at Fraunhofer IBMT, follows a similar pattern, combining a modular phased-array front end with an integrated PC and GPU, supporting the real-time processing of multi-channel data, and providing open interfaces for custom beamforming and imaging algorithms~\cite{risserHighChannelCount2016,boniUltrasoundOpenPlatforms2018}.

\paragraph*{Commercial programmable research systems}

Commercial programmable systems, such as the Verasonics Vantage, the US4R platform, and the Open Sonics system, also target research applications~\cite{verasonicsVantageNXTResearch,us4usUs4R,opensonicsUltrasoundResearchProducts}. These systems offer high channel counts, configurable transmit-receive sequences, and access to raw radiofrequency (RF) data for advanced processing on host workstations.
These systems have become the de facto standard in many experimental ultrasound laboratories. However, they remain proprietary at the hardware and firmware levels: users can script high-level imaging sequences and implement custom reconstruction or postprocessing in software, but they cannot modify the internal signal chain, low-level timing, or embedded real-time processing blocks. 

\paragraph*{Compact and low-cost modular platforms}

Alongside high-end open scanners, recent work has explored compact, low-cost, and modular research platforms. A relevant example is the SoC-based design by Tageldeen et al.~\cite{tageldeenCompactModularOpen2025}, which emphasizes software-centric imaging pipelines with interchangeable analog front-end modules and on-board processing on embedded CPUs and GPUs. These systems offer extensive software control over transmit parameters, receive paths, and reconstruction pipelines, making them particularly suitable for rapid prototyping and teaching. However, current prototypes are typically limited in channel count and supported imaging modes compared to large open scanners or commercial research systems.

\subsection{Summary and key limitations}

Several common architectural patterns are present on both open and commercial platforms~\cite{boniUltrasoundOpenPlatforms2018}. 
First, most systems heavily rely on FPGAs for core tasks such as high-speed data acquisition, transport, control, and beamforming. Although FPGAs provide the necessary throughput to interface with modern ultrasound front-ends, this dependence often restricts low-level reconfigurability to experts in HDL programming~\cite{kimSingleFPGAbasedPortable2012,wuFPGABasedBackendSystem2019}. 

Second, nearly all of these systems share a similar strategy for handling high-bandwidth data: groups of up to 32--64 RX/TX channels are managed by an FPGA-based front-end module, which is connected to local RAM buffers. High-rate raw acquisition data are first written into these buffers and then transferred in bursts to a workstation or on-board PC via PCI Express, Ethernet, or proprietary links. Although local buffering increases robustness and decouples acquisition from host-side jitter, it limits the duration of high-data-rate acquisitions and complicates continuous real-time streaming at very high frame rates. Furthermore, despite the fact that new data transmission protocols pervasive in high-performance computing systems, such as Remote Direct Memory Access (RDMA), hold great promise for improved transmission bandwidth and duration \cite{guoRDMACommodityEthernet2016}, they remain largely underexplored for ultrasound systems~\cite{cossettiniRDMAInterfaceUltraFast2022}.

Third, time-sensitive operations, such as triggering and sequencing transmit events, as well as multi-channel digitization, are typically implemented as state machines within the FPGA fabric. Even when the analog and digital front ends are software-configurable, the host executes configuration loops that update the FPGA's registers and control memories before acquisition. Consequently, making fine-grained modifications to acquisition behavior during ongoing scans is a challenging task. Changing beam sequences, timing relationships, or channel routing typically requires reprogramming embedded state machines, re-synthesizing FPGA logic, or interrupting the acquisition to reload configuration memories.

In summary, existing platforms span a wide spectrum, ranging from powerful but closed commercial systems with high-level scripting capabilities to FPGA-based open systems that still demand low-level programming for significant architectural changes. In addition, there are compact smart probes with fully software-defined pipelines but limited channel counts, as well as closed consumer devices with no research access. What remains absent is a fully open, high-channel-count ultrasound system in which signal acquisition, processing, and image reconstruction are defined entirely in software, while dedicated hardware is restricted to pulsing, amplification, digitization, and streaming.

The LtL platform proposed in this work addresses these gaps by exploiting the latest-generation of AFEs with sampling frequencies up to 125 MSPS, a Zynq UltraScale+ MPSoC with tightly-coupled FPGA and Application Processing Unit (APU) for software-defined operation, and RDMA streaming for increased datarate (as high as 95.4 Gbit/s from a single FPGA and network card). In particular, all high-level imaging behavior (including triggering, sequencing, and data handling) is configured at run time in software on the APU. This paper validates the key building blocks of LtL in a 16-channel prototype and quantitatively compares LtL to SoA.

\section{System Concept}
\label{sect:system_concept}

\begin{figure*}
    \centering
    \includegraphics[width=\textwidth]{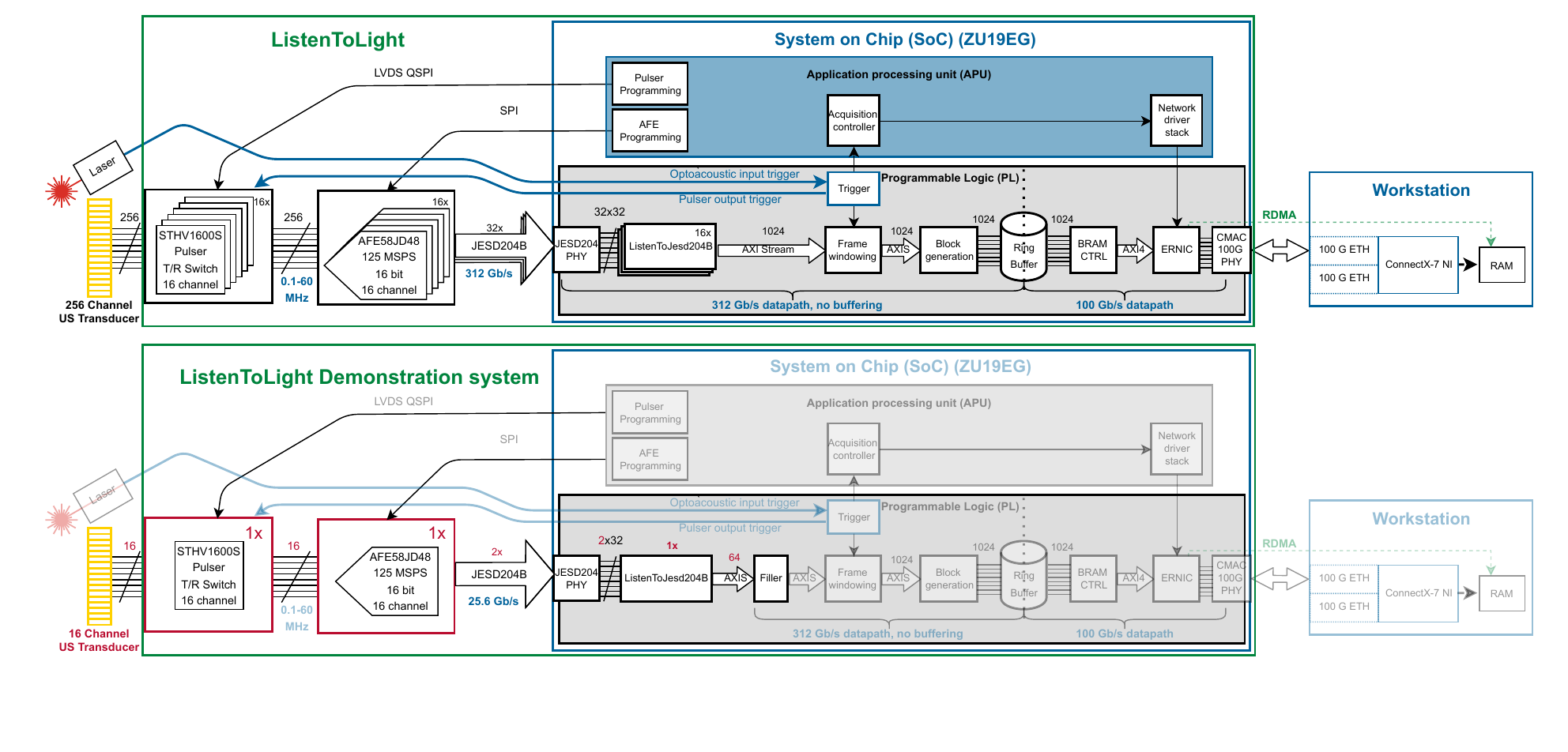}
    \caption{Top: Overview of the LtL system architecture. Bottom: 16-channel demonstration implementation (red boxes indicate the differences between the 256-channel system concept and the 16-channel validation).}
    \label{fig:system_design}
\end{figure*}

The ListenToLight (LtL) platform is conceived as a scalable, 256-channel, software-defined, dual-mode (OA/US) ultrafast imaging system.
To enable ultrafast imaging in this shared mode, LtL must support high-bandwidth processing, long acquisition periods, and a flexible, runtime-programmable configuration.
Fig.~\ref{fig:system_design} provides an overview of the complete system architecture.

LtL integrates 256 transmit/receive (TX/RX) channels in a full-matrix configuration. For TX, LtL features sixteen STHV1600 high-voltage pulsers (16-channel, 5-levels, 200Vpp/4 A per channel up to 50 MHz, 5ns resolution, integrated T/R switches; STMicroelectronics).
We selected this component as a compromise between software controllability, channel density, and drive strength.
The pulse is controlled by software, and the output waveform is determined by its internal memory, which can store up to 256 transmission patterns. Additionally, each channel can be individually controlled with a resolution as fine as \qty{5}{ns}.

For RX, LtL features sixteen AFE58JD48 analog front-end (AFE) devices (16-bit, 16-channel, 125 MSPS ADC, -3 dB to 48 dB gain range; Texas Instruments, USA). At the time of writing, this is the most advanced multi-channel AFE for ultrasound applications commercially available. Its channel grouping (8 channels per single JESD204B link) and its integrated digital processing blocks (filtering, decimation, and I/Q demodulation) allow the acquisition of RF data at 16~bit, 80~MSPS per channel using only two high-speed interfaces JESD204B, or decimated/demodulated data at 125 MSPS when the onboard DSP is used. Additionally, it integrates a programmable active termination that allows for active termination of the RF line (preferred for conventional ultrasound) or maintaining it at high impedance (usually preferred for optoacoustic ultrasound).  
The combination of AFE and pulser and their wide range of operation enables the use of LtL for all commercially available and medically relevant transducers, as long as their frequency bandwidth lies between the AFEs input 10 kHz highpass and 62.5 MHz Nyquist limit for the 125~Msamples/s (MSPS) sampling frequency, and the pulser's 50 MHz maximum output frequency. 

LtL is built around a single Multi-Processor System-on-Chip (MPSoC), the Zynq Ultrascale+ ZU19EG (Advanced Micro Devices, USA), integrated in the iW-G35M System-on-Module (SoM) by iWave Systems Technologies (India). Using an MPSoC allows software control by an APU to be tightly integrated with the high-bandwidth datapath of the FPGA. Within the MPSoC EG Devices, the ZU19EG has been specifically chosen for the memory and transceiver capabilities, as it features 32 16.3 Gb/s GTH and 16 32.75 Gb/s GTY transducers, the first group supporting up to 256 AFE channels in total, the second up to 4 100 Gbit ethernet physical layers. Data communication to a workstation is implemented via an optical link for 100 Gigabit Ethernet (100GbE).

Using a standard, low-cost MPSoC enables the system to run with a tight and low component count highly integrated hardware architecture. Crucially for an ultrasound system, the ZU19EG integrates an application processing unit in the form of an Arm Cortex-A53 quad-core processor and a programmable logic (PL) fabric, as well as DDR Memory interfaces. Having a tightly integrated processor and fabric is fundamental to the novel concept of ``buffer-free'' ultrasound acquisition. This tight coupling reduces latencies, enables software-defined sequencing, and guarantees real-time software-based management of RDMA triggers. 

\subsection{Data and clocking interfaces}
\label{sec:data-clocking}

To operate 256 TX/RX channels with a monolithic system, we selected JESD204B as the data interface between the AFE and the PL. This allows for the reduction of the number of physical data lines required for data transfer while maintaining high data rates. Each AFE58JD48 device can transmit the output of 8 ADC channels over a single JESD204B link (more details in Sect. \ref{sect:AFE_16_data}). Thus, 32 links are sufficient to handle all 256 receive channels of the system.

To enable ultrafast data transmission to a workstation without being limited by any OS-mediated network stack, we implement a Remote Direct Memory Access (RDMA) protocol over the 100GbE link. 

In multichannel ultrasound systems, synchronizing clocks and samples is essential because all received signals must be aligned in time for accurate beamforming. In our design, this means ensuring that every analog-to-digital converter (ADC) across all 16 analog front-end (AFE) devices samples data simultaneously and that the data can be perfectly aligned inside the FPGA, despite differences in transmission delays.

To achieve this, all ADCs are driven by a shared, low-jitter external clock that generates the AFE sampling clock (AFECLK), which defines the acquisition time base for all 256 channels. By distributing a single AFECLK signal, we ensure that every AFE samples data simultaneously. The AFECLK is fanned out through matched clock buffers and length-matched PCB traces to each AFE and to the FPGA, minimizing inter-device sampling skew.

At the transport level, each AFE sends its digitized samples to the FPGA using JESD204B Subclass 1 links for data transfer. Deterministic latency is achieved by means of a system reference signal (SYSREF), derived from the same external clock source. SYSREF is generated at a parameter-dependent submultiple of AFECLK and is distributed with a matched delay to all AFEs and to the FPGA. Each AFE captures this signal using AFECLK and uses it to align the start of its JESD data frames with a specific ADC sample. Since all AFEs are configured the same way, they behave identically when handling data.

On the FPGA side, the SYSREF signal enables the JESD204B receiver logic to precisely align all incoming data streams. After deserialization and decoding, samples originating from different AFEs but corresponding to the same acquisition instant are aligned by ListenToJESD204B elastic buffer in the FPGA fabric, regardless of JESD lane skew or link-dependent latency differences.

This top-down synchronization strategy, which utilizes a single clock source through deterministic JESD framing, ensures that all channels stay perfectly aligned within the FPGA. This precise alignment enables coherent beamforming and high-quality ultrasound signal processing.

\subsection{Processing and control platform}

The ZU19EG MPSoC enables a compact and tightly integrated system architecture by combining processing, control, and high-throughput data handling within a single device. In contrast to conventional ultrasound backends that rely on multiple discrete processing and control components, this monolithic architecture combines typically separate functions within a single MPSoC, thereby reducing system complexity and inter-component latency. Crucially for an ultrasound system, the ZU19EG integrates both an application processing unit (APU), implemented as a quad-core Arm Cortex-A53 processor, and a large PL fabric, together with high-bandwidth DDR memory interfaces. This tight coupling between processor and fabric is fundamental to the proposed "buffer-free" ultrasound acquisition concept, as it enables low-latency interaction between software and hardware, software-defined sequencing, and real-time software-based management of RDMA transfers.

Within the ZU19EG, responsibilities are cleanly partitioned between the MPSoC application processing unit (APU) and the FPGA PL. The APU, running a Linux operating system, configures the AFEs and pulsers via Serial Peripheral Interface (SPI) and Quad-SPI (QSPI), manages acquisition parameters, and controls the network interfaces to orchestrate data transmission to the workstation. Time-critical operations are implemented in the PL, which receives digitized data from the AFEs over JESD204B links and performs frame windowing, buffering, and high-throughput data transfer to the host PC via a 100~GbE interface using RDMA.

With this organization, ListenToLight can be regarded as a fully software-defined ultrasound platform, in which all system parameters are configured from the Linux environment running on the SoC's Arm cores, while the PL implements a deterministic, low-latency datapath. The PL therefore acts primarily as a streaming pass-through, following a fixed chain of JESD204B acquisition, lightweight buffering, and RDMA transmission, as described in more detail in Sect.~\ref{sect:FPGA_PL_datapath}.

\section{Demonstration Implementation (16-channel prototype)}
\label{sect:sixteen}

We conceive the 256-channel architecture as a modular solution, with multiple modules sharing a global reference clock, a common trigger, and a common configuration interface from the APU.
To validate the LTL architecture, we implemented a 16-channel demonstration platform using commercially available evaluation boards. This prototype preserves all the key architectural components envisioned for the full 256-channel system, including clocking strategy, ultrafast JESD204B data transmission between AFE and FPGA, 100 GbE streaming to a host workstation, and software/firmware stack.
In the following, although only 16 channels are digitized, the PL datapath always processes 256-channel-wide AXI streams with 240 dummy channels inserted to evaluate realistic throughput.

Fig.~\ref{fig:system_design} summarizes the demonstration prototype, as detailed in Sect.~\ref{sect:architecture16}.

\subsection{Hardware Architecture}
\label{sect:architecture16}

The demonstration prototype consists of the following components:
\begin{itemize}
    \item The iW-G35M SoM (featuring the ZU19EG MPSoC), mounted on its Development Carrier board (iW-G35D);
    \item One AFE58JD48EVM evaluation board, featuring one 16-channel AFE with 16-bit resolution and up to 125 MSPS sampling rate;
    \item One STHV1600EVM evaluation board, for 16-channel pulsing and T/R switch functionality;
    \item A workstation with a Mellanox Connect-X 7 network interface controller (NIC);
    \item A 32-channel, 5 MHz linear array transducer (LA5.0/32-2077, Vermon, France), shorting neighboring elements to obtain 16 effective channels;
    \item A laser diode (LD)-based OA excitation source (QD-Qxy10-ILO, Quantel laser, France), emitting pulsed light at a wavelength of 808 nm, with a pulse energy of 1 mJ and a pulse width of 100 ns.
\end{itemize}

The SoM is connected to the AFE evaluation board via an FMC connector and to the workstation via a QSFP cage. 

\subsection{Clocking and Synchronization}

The AFE EVM board supplies a SYSREF and ADCCLK to both AFE and PL, as described in section \ref{sec:data-clocking}. For the demonstration system, the ADCCLK towards the AFE has been set to 80 MHz, and ADCCLK towards the FPGA has been set to 320 MHz (equivalent to 1/40th of the 12.8Gbps line rate). The SYSREF clock has been set to 2.5 MHz. The SYNC signal needed to start the Initial Lane Alignment sequence is supplied by the ListenToJESD204B block. 

Within the SoM, the APU (see Sect.~\ref{sect:APU_sw}) provides the internal logic clock source for the acquisition and network interfaces, and connects to the PL in two main ways: through an AXI-4 interconnect to control the ERNIC and PL peripherals, and through an interrupt bus to receive the interrupts from the PL indicating new frame acquisitions. 

\subsection{AFE data output}
\label{sect:AFE_16_data}

Each 16-channel AFE58JD48 transmits data over two JESD204B links (one for each 8-channel ADC bank), at up to 12.8 Gb/s.
Two configurations can be selected for data transmission: raw data or I/Q demodulation.

In raw data transmission mode, data from 8 channels is transmitted over a single JESD204B link at 12.8 Gb/s. Since the JESD204B link is 8b/10b encoded, this results in an effective data rate of 10.24 Gb/s and a maximum acquisition speed of 80 Msamples/s per channel.

When implementing data reduction (with I/Q demodulation and decimation) on the AFE, each 8-channel JESD204B link can transmit data, centered around the transducer center frequency, at a rate of 10 Gb/s per link. In this configuration, a maximum sampling rate of 125 Msamples/s can be achieved per channel. 

\subsection{FPGA (Programmable Logic) datapath design}
\label{sect:FPGA_PL_datapath}

The PL is organized into three main functional blocks: a JESD204B frontend for data reception and alignment, a buffering and packetizer block, and an Embedded RDMA-enabled NIC (ERNIC).

\subsubsection{JESD204B link} 

Data from each JESD204B link group (2 links per AFE) is received on the FPGA by a JESD204B RX PHY IP core (AMD, USA), which deserializes the incoming data and presents a parallel 32-line interface (at 4$\times$ the sample frequency for raw-data-mode and 2$\times$ the sample frequency for I/Q-data-mode). 

\begin{figure}
    \centering
    \includegraphics[width=0.95\columnwidth]{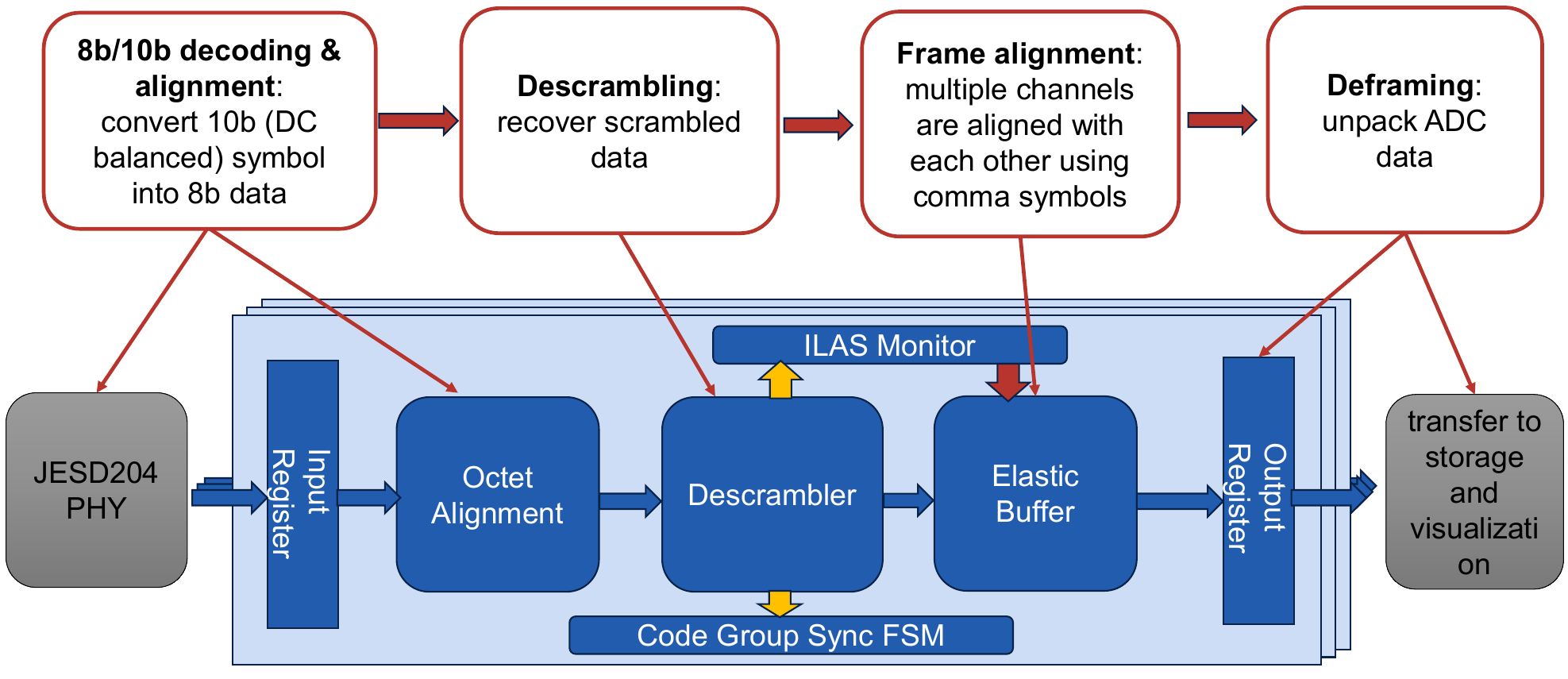}
    \caption{JESD implementation (adapted from \cite{bhattacharjeeListenToJESD204BLightweightOpenSource2025}).}
    \label{fig:JESD_implementation}
\end{figure}

Next, an open-source JESD IP core (ListenToJESD204B, \cite{bhattacharjeeListenToJESD204BLightweightOpenSource2025}), shown in Fig.~\ref{fig:JESD_implementation}, performs Code Group Synchronization (CGS), validates initial lane assignment, performs optional descrambling, and aligns the data between the two lanes in the device using elastic buffers per-lane. The presence of elastic buffers enables the absorption of small differences in lane delays, allowing the use of non-matched high-speed JESD204B traces. 

At the exit of the ListenToJESD204B, the data is presented as an AXI-Stream interface with signals indicating the start of frame and error conditions. The Sysref signals of standard Subclass 1 JESD204B operation are used by the AFE to set the start of multiframe alignment. 
The multiframe alignment signal is thus synchronized with the same sample for all possible AFEs in the system, regardless of the individual device lane delays. This condition ensures that we can extend the 16-channel architecture to a future 256-channel system without requiring changes to the interfaces. 

After this initial JESD204B processing, the data is passed through a clock domain crossing FIFO into the single PL system clock domain. When connecting multiple AFEs, alignment is ensured by delaying all individual streams to their respective multiframe start signals and coalescing them into a single 1024-bit-wide AXI-Stream interface.

Such a wide AXI-Stream represents the 256-channel datapath. For the sake of the 16-channel validation presented in this paper, unused channels are padded with dummy data to emulate the full system data width.

\subsubsection{Framing and Buffering}

Since the JESD stream is continuous, the PL window is used to generate individual frames. Windowing is performed based on a trigger signal (that originates from either the external laser trigger or internally, triggering the pulser) and is handled by the Trigger block of Fig. \ref{fig:system_design}. Running the trigger on the FPGA guarantees a precise temporal alignment of the acquisition to the laser firing or pulser output. The input trigger is latched by the AFE sampling clock, resulting in a maximum capture latency of $ 1/\qty{80}{MHz} = 12.5ns$.
Once a trigger event is detected, a programmable delay is applied before data capture begins to accommodate system-specific timing requirements. After this delay expires, the Trigger block activates the Frame Windowing block. This block forwards a programmable number of samples to the buffering logic while simultaneously asserting an interrupt to the APU, signaling the start of a new frame acquisition.

The Block Generator then writes the incoming data into a circular ring buffer implemented as a simple dual-port URAM array. Data are stored in contiguous blocks of programmable size at consecutive physical memory locations, enabling efficient burst transfers. The buffer supports concurrent write operations for frame acquisition and read operations for RDMA-based data transfer, as described in the following section.

\subsubsection{RDMA}

After the first block is filled, the APU orchestrates the RDMA transfer of all packets that are scheduled to be written to the circular buffer to the host PC memory (see Sect.~\ref{sect:APU_sw}). Then, the ERNIC IP core begins reading data from the buffer over a memory-mapped AXI-BRAM interface, utilizing the internal DMA engine while the acquisition logic writes data.
As the datapath of the ERNIC core is slower than the incoming data rate from the frontends, the only limiting factor of each frame size is given by the ``leaky bucket" relationship between writing and reading speeds (see Sect.~\ref{sec:frame_size_fps} for a more detailed characterization). 

\subsection{APU software design}
\label{sect:APU_sw}

The APU runs a custom Petalinux 2023.2 distribution, extended with drivers for the ERNIC, custom \texttt{uio} interfaces for PL interrupts, and SPI/QSPI drivers for configuring the AFE and pulser. A user-space application orchestrates the configuration of the frontends, manages acquisition parameters, and handles data transfers between the PL and the host PC via RDMA.

\subsubsection{RDMA initialization}

The RDMA communication runs from user space. 
Before the RDMA data transfer can begin, the APU establishes an RDMA connection to the workstation using the RDMA CM protocol. This process involves the following steps:
\begin{itemize}
    \item Connection handshake: the SoC and workstation create queue pairs (QPs) to exchange CM messages via Unreliable Datagram (UD);
    \item After the above handshake, both RDMA endpoints transition to Reliable Connection (RC) QP to perform data operations (RDMA WRITE, SEND);
    \item The workstation performs an RDMA SEND operation to transmit its memory registration information to the APU. This allows RDMA WRITE operations (issued by the SoC) to directly transfer data to the registered input buffer located in the workstation's main memory. 
\end{itemize}

\subsubsection{RDMA operation}

Synchronization between the data producer (JESD204B interface), user-space application, and ERNIC is achieved through an interrupt-driven mechanism implemented via a custom \texttt{uio} driver. The driver maintains an internal interrupt counter that records the number of interrupts generated since the last \texttt{read()} system call. Each interrupt indicates that a 256\,KB block in the ring buffer has been filled by the producer.

The LTL user-space application periodically reads from the \texttt{/dev/uio} device to determine the number of interrupts (and thus the number of new data blocks) that have occurred since the last check. For each newly filled block, the application posts a corresponding RDMA WRITE WR to the ERNIC, specifying the address of the block in BRAM as the data source. The ERNIC then performs the RDMA WRITE operation, transferring the data directly into the destination buffer on the workstation.

\subsubsection{RDMA data transmission} 

\begin{figure}
    \centering
    \includegraphics[width=0.9\columnwidth]{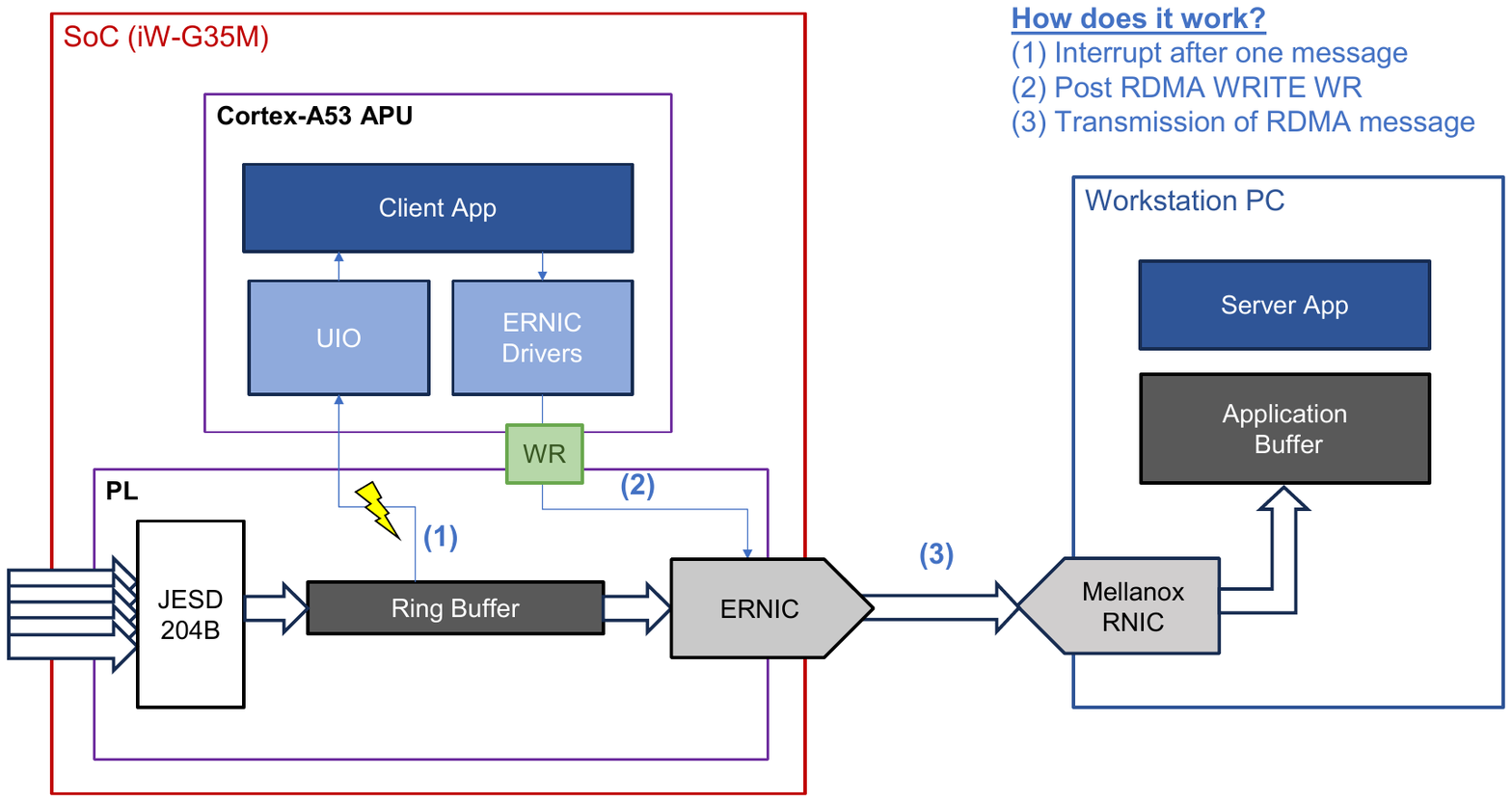}
    \caption{Overview of 100G RDMA interface architecture}
    \label{fig:overview_rdma_arch}
\end{figure}

After initializing QPs and registering the workstation memory, data transmission proceeds as follows (see also Fig.~\ref{fig:overview_rdma_arch}):

\begin{itemize}
    \item The user-space application polls the frame generator interrupt line, indicating a ``frame start'' event;
    \item As soon as an interrupt is detected, the user-space application posts a batch of RDMA WRITE WR to the ERNIC, specifying the addresses of the block in the BRAM as the data source;
    \item The ERNIC then performs the RDMA WRITE operations autonomously, transferring the data directly into the destination buffer on the workstation;
    \item The user-space application detects that the transmission is completed by polling the completion queue (CQ). 
\end{itemize}

In the demonstration system, a ``single-frame" mode was used to characterize the maximum transmission rate of the ERNIC block by triggering the transmission of a new WR Batch immediately after a transmission complete is detected. These measurements are shown in Sect.~\ref{sec:frame_size_fps} and show the maximum achievable transmission rate for this acquisition scheme, confirming the available bandwidth to support 256 channels.

\subsection{Acquisition modes}

The demonstration platforms support the same imaging modes as the 256-channel concept architecture. 

\textbf{US pulse-echo imaging.} The acquisition flow to demonstrate the US pulse-echo capabilities of our platform is as follows:
\begin{enumerate}
    \item The host PC configures the pulser via the APU;
    \item The host PC sends a ``start acquisition'' command;
    \item The APU starts the pulser after a programmable delay;
    \item The PL circular buffer starts to capture data from the frontends\footnote{\label{footnote:AFE}the frontends stream continuously};
    \item As soon as one of the packets in the circular buffer is filled, the RDMA driver in the APU orchestrates the transfer of the data to the host PC and starts a batch transfer immediately;
    \item Data is consumed from the circular buffer while new data is being written to the next packet;
    \item The data is received on the host PC memory and available for further processing.
\end{enumerate}

\textbf{OA imaging.} The acquisition flow to demonstrate the OA imaging capabilities of our platform, when used with an external laser source, is as follows:
\begin{enumerate}
    \item A function generator triggers simultaneously the external laser and the PL;
    \item After a programmable delay, a PL circular buffer starts to capture data from the frontends;\footnotemark[1]
    \item the RDMA data transfer and consumption, then proceed as in the US pulse-echo case.
\end{enumerate}

\section{Hardware characterization}
\label{sect:characterization}

\subsection{Receive path bandwidth and SNR}

The frequency response of the RX chain was characterized at \qty{125}{MSPS} by exciting the ADC input with a single-tone sine wave from a signal generator (constant \qty{100}{mVpp} source setting into \qty{50}{\ohm}) and sweeping the tone frequency across the usable band (up to \qty{62.5}{MHz}). For each tone, a contiguous record of 32768 raw ADC samples was captured and processed directly in the frequency domain using a single FFT per tone. 

\paragraph{Bandwidth}
Bandwidth was derived from the measured time-domain amplitude of each captured record: for each tone frequency, the peak-to-peak code excursion $A_{\mathrm{pp}}(f)$ was computed and normalized by the maximum peak-to-peak amplitude over the sweep. The gain in dB was then computed as
\begin{equation}
G_{\mathrm{dB}}(f) = 20\log_{10}\!\left(\frac{A_{\mathrm{pp}}(f)}{\max_{f} A_{\mathrm{pp}}(f)}\right).
\end{equation}
The lower and upper \SI{-3}{dB} corner frequencies were then obtained by locating the first crossings of $G_{\mathrm{dB}}(f)=-3$ on each side of the peak-gain frequency and linearly interpolating between adjacent sweep points.

\paragraph{SNR}
For each tone, the sample mean was removed and a one-sided FFT was computed. Signal power $P_{\mathrm{sig}}$ was obtained by integrating the FFT-bin power in a neighborhood around the fundamental bin, with a span set to the larger of (i) a fixed bin span and (ii) a small fraction of the fundamental bin index (to keep the mask proportional at higher frequencies). Noise power $P_{\mathrm{noise}}$ was obtained by integrating FFT-bin power from \qty{1}{MHz} to Nyquist while excluding DC, the fundamental region, and guard regions around the first few harmonics and subharmonics (to prevent tone leakage or deterministic spurs from inflating the noise estimate). 

The in-band SNR was then computed as
\begin{equation}
\mathrm{SNR}_{\mathrm{dB}} = 10 \log_{10}\!\left(\frac{P_{\mathrm{sig}}}{P_{\mathrm{noise}}}\right),
\end{equation}

The measurement of bandwidth and SNR revealed a \qty{-3}{dB} RX bandwidth from \qty{0.99}{MHz} to \qty{46.35}{MHz}, referenced to the peak-gain point (at \qty{8.00}{MHz}). The magnitude response rises from about \qty{-8.9}{dB} at \qty{0.2}{MHz} to within \qty{0.25}{dB} of peak gain by \qty{4}{MHz}, remains approximately flat (within \qty{0.25}{dB}) through \qty{15}{MHz}, and then rolls off smoothly toward Nyquist (about \qty{-9.4}{dB} at \qty{62.49}{MHz}). The full-scale-normalized SNR is over \qty{56}{dB} for the entire measured range, thus signals outside the \qty{-3}{dB} bandwidth can still be acquired with high fidelity.

\begin{figure}[t]
    \centering
    \begin{subfigure}[b]{0.9\columnwidth}
        \centering
        \includegraphics[width=\linewidth]{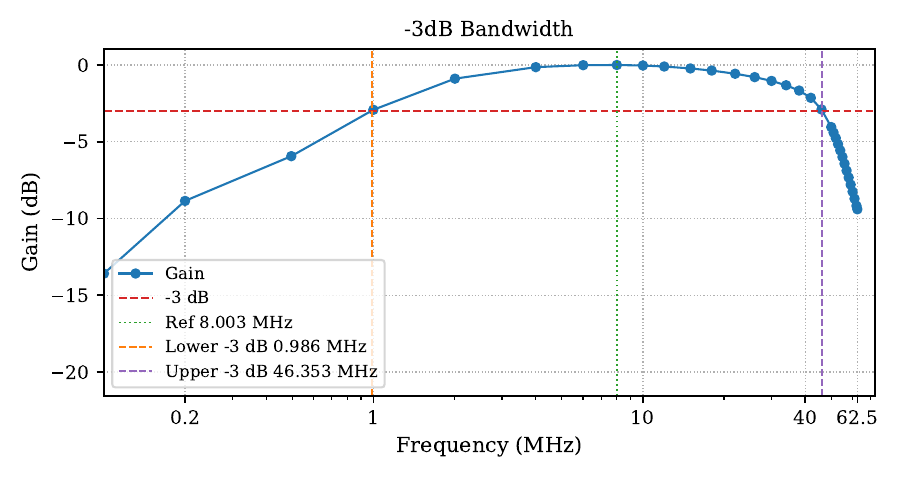}
        \caption{Measured RX magnitude response and \qty{-3}{dB} bandwidth using a swept sine stimulus.}
        \label{fig:rx_bw_snr_mag}
    \end{subfigure}

    \medskip

    \begin{subfigure}[b]{0.9\columnwidth}
        \centering
        \includegraphics[width=\linewidth]{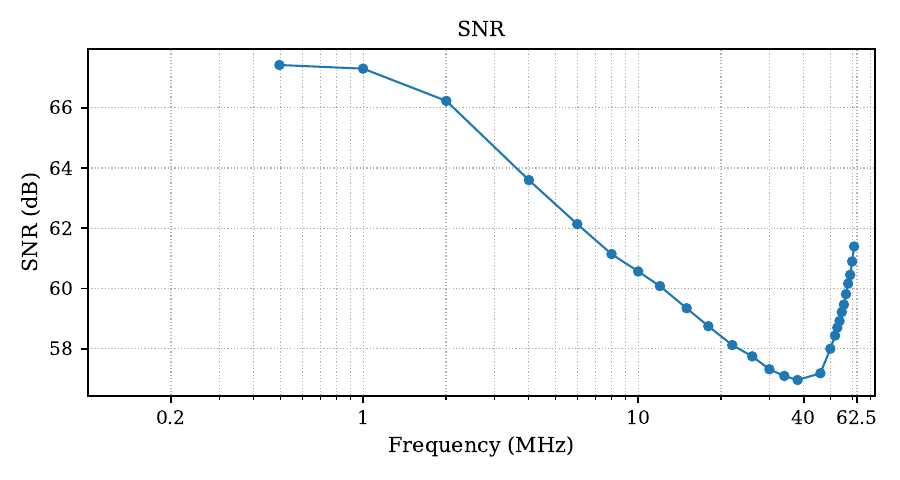}
        \caption{SNR versus input frequency.}
        \label{fig:rx_bw_snr_masks}
    \end{subfigure}

    \caption{RX characterization.}
    \label{fig:rx_bw_snr}
\end{figure}

Fig.~\ref{fig:rx_bw_snr} summarizes the measured RX magnitude response and the SNR across the ADC measurement range.

\subsection{Frame size and frame rate limit}
\label{sec:frame_size_fps}

During acquisition, the PL writes samples in the circular buffer, while the ERNIC concurrently reads previous blocks after an initialization delay.
Since the buffer is circular, the maximum frame length is not limited by the buffer size, but by the ``leaky bucket" relationship between the write speed and read speed. We can express the relationship as: 

\begin{equation}
L_{\text{f,max}}
= R_{\text{in}}^{\text{eff}} \,
  \frac{B - R_{\text{out}}^{\text{eff}} \, \tau_{s}^{\text{eff}}}
       {R_{\text{in}}^{\text{eff}} - R_{\text{out}}^{\text{eff}}},
\end{equation}

where  $L_{\text{f,max}}$ is the maximum allowed frame length (expressed in bits) that can be safely written into the buffer without causing overflow under worst-case conditions. The term $R_{\text{in}}$ represents the buffer write rate. $R_{\text{out}}^{\text{eff}}$ is the effective worst-case buffer read rate (taking into account transmission speed deviations, such as retransmission). The parameter $B$ is the usable buffer capacity, expressed in bits, and $\tau_{s}^{\text{eff}}$ represents the worst-case read-start latency. 

For a 4 MiB buffer, write rate $R_{\text{in}} =$ 327 Gbit/s (corresponding to 256 channels at at 16 bit, 80 MSPS sampling), and measured ERNIC read speed of 95.6 Gbit/s (see Sect.~\ref{sect:results_speed}), the resulting maximum frame length is 5639 samples per channel, and the maximum sampling frequency is $R_{\text{out}}^{\text{eff}}/L_{\text{f,max}} = FPS_{max}$ 4.14 kFPS (not considering pulser T/R duration). 

The maximum frame acquisition speed and frame length for LtL can be derived from the above formulas for different acquisition sampling speeds, channel counts, and acquisition duration. For example, using a conventional UFUS sample number such as 2000 samples per frame, the achievable framerate is 11.7 kFPS.
Similarly, by reducing the sampling rate to 36 MSPS and 10-bit, similar to other related works, or reducing the active channel count, full streaming of uninterrupted ADC data is possible (see also the discussion in Sect.~\ref{sect:soa}).

Compared to prior works, this approach is not limited by the duration of the acquisition, but only by the frame length. In fact, the overall length of the acquisition is only limited by the workstation's memory or high-bandwidth storage. 

\subsection{RDMA link throughput measurement}
\label{sect:results_speed}

The maximum FPGA-to-PC data throughput over RDMA was measured by configuring LtL to continuously stream dummy data from the BRAM circular buffer to the PC Memory, simulating a new transmission request sent immediately after the previous one is completed (no JESD connection is needed for this stress test). Payload sizes ranging from 64 KB to 1 MB within Batch transfers ranging from 1 to 16 were used to simulate different acquisition scenarios.\footnote{In this context, the batch size refers to the number of RDMA WRITE WRs posted to the SQ at once before polling the CQ.} Throughput measurements were averaged across 10 test runs.

\begin{figure}[ht]
    \centering
    \includegraphics[width=0.8\columnwidth]{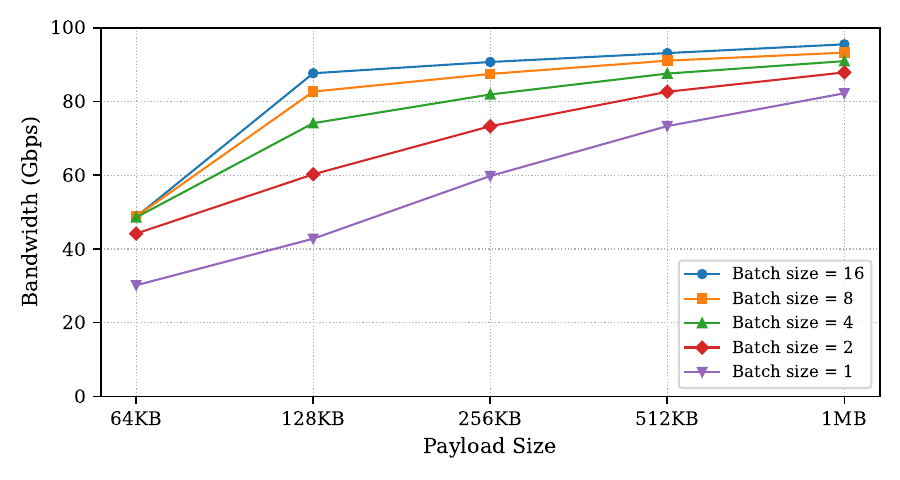}
    \caption{Bandwidth vs payload characterization of RDMA transfers.}
    \label{fig:data_troughput}
\end{figure}

Fig.~\ref{fig:data_troughput} shows the corresponding bandwidth measurements. 
As expected, data throughput increases with both payload size and batch size, reaching a maximum of 95.6 Gbps for a payload size of 1 MB and a batch size of 16. For the payload and batch size of the validation measurements reported in the next section (256KB and 8, respectively), the measured maximum throughput was 87 Gbps.
Inter-test variability remained within $\pm$ 0.1 Gbps.

\subsection{Resource utilization and Power consumption}

Table \ref{tab:power_consumption} reports the power consumption of the demonstration system. The power for the AFE and pulser evaluation boards was measured with an N6705B power analyzer (Keysight Technologies, USA). PL and PS power was estimated using the AMD Vivado post-implementation Power consumption estimation. 

\begin{table}[ht]
    \centering
    \caption{Power consumption of the main components of LtL}
    \label{tab:power_consumption}
    \small
    \begin{tabular}{@{}lcc@{}}
    \toprule
        \multicolumn{2}{@{}l}{\textbf{component}} & \textbf{power consumption} \\
    \midrule
        \multicolumn{2}{@{}l}{AFE*} & 6.7 W  \\
        \multirow{2}{*}{Pulser*} 
            & Low Voltage & 1.19 W  \\
            & High Voltage & 30 mW (10V), 20 mW (-10V) \\
    \midrule
        \multirow{4}{*}{FPGA}
            & JESD204B& \qty{7.58}{W} \\
            & APU& \qty{3.35}{W} \\
            & 100G Subsystem& \qty{4.42}{W} \\
            & Interconnect& \qty{1.479}{W} \\
    \bottomrule
    \end{tabular}\\[0.5ex]
    \footnotesize\textsuperscript{*}Evaluation Boards
\end{table}

Table \ref{tab:fpga-utilization} summarizes the FPGA resource utilization for the 16-channel demonstration system.
Towards the 256-channel platform, only the JESD204B acquisition blocks scale linearly with the channel count, whereas the remainder of the datapath is already synthesized for 256-channel width, allowing for the reuse of the same backend for the final prototype with minimal increase in FPGA resources.

Overall resource utilization is low (e.g., below 20\% of LUTs), leaving substantial headroom for future expansions. Possible expansions include integrating additional on-board processing blocks, connecting up to 96 more AFE channels, or instantiating up to three additional 100 GbE subsystems using currently unused GTY ports, or increasing the on-chip RF frame Buffer to support larger maximum frame sizes.

The monolithic system architecture, based on a single MPSoC, combined with the low resource utilization, enables the processing of the entire streamed data set onboard without requiring inter-chip communication. Relying on a single processing device for acquisition, control, and high-throughput data handling distinguishes this architecture from existing state-of-the-art platforms in the same class.

\begin{table}[h]
  \centering
  \caption{ZU19EG PL resource utilization by subsystem}
  \label{tab:fpga-utilization}

  \setlength{\tabcolsep}{1.5pt}
  \begin{tabular}{@{}llccc@{}}
    \toprule
    Subsys. & Variant & FF & LUT & BRAM*  \\
    \midrule
    \multirow{2}{*}{JESD} 
            & 16\,ch  & 1.4k (0.2\%) & 0.9k (0.2\%) & 0  \\
            & 256\,ch & 8.1k (0.8\%) & 5.1k (1\%)   & 0  \\
    \midrule
    \multicolumn{2}{@{}l}{100G subsys.} & 68.8k (6.6\%) & 48.2k (9.2\%) & 1.5 (17\%)  \\
    \multicolumn{2}{@{}l}{RF buffer}    & 3k (0.3\%)    & 1.3k (0.3\%)  & 2.2 (25\%)  \\
    \multicolumn{2}{@{}l}{Interconnect} & 44k (4.2\%)   & 29.1k (5.6\%) & 0   \\
    \bottomrule
  \end{tabular}\\[0.5ex]
  \scriptsize\textsuperscript{*}BRAM in MB. DSP: 2 (0.1\%)
\end{table}

\section{Validation Phantom Measurements}
\label{sect:validation_phantom}

This section reports validation measurements on US and OA phantoms. The focus in this work is not on clinical validation and imaging quality, but on validating end-to-end data acquisition, timing alignment, and RDMA streaming.

\definecolor{darkgreen}{RGB}{0,120,0}

\begin{table*}[!ht]
  \centering
  \scriptsize
  \caption{Comparison of 256 Channel research ultrasound and hybrid OA/US imaging platforms (offering raw data access)}
  \label{tab:platform-comparison}
  \begin{tabular}{@{}lcccccc@{}}
    \toprule
    Device & Modality & Pulser / TX & AFE / ADC\footnotemark[1] & Buffer\footnotemark[2] & Streaming\footnotemark[3] & Peak\footnotemark[4] [GB/s] \\
    \midrule
    ULA-OP~256 \cite{boniULAOP256256Channel2016} & US & \textcolor{darkgreen}{\textbf{20MHz AWG 200Vpp}} & 12-bit @78.1 & \textcolor{darkgreen}{\textbf{144~GB}} & N/S (USB 3.0) & 0.2 \\
    DiPhAS \cite{risserHighChannelCount2016} & US & 3-lvl 30MHz 200Vpp & 12-bit @80 & 64~GB* & N/S (PCIe DMA) & $\approx$1.5 \\
    Verasonics NXT \cite{verasonicsVantageNXTResearch} & US & 3-lvl 60MHz, 2--190Vpp & \textcolor{darkgreen}{\textbf{16-bit @125}} & 64~GB & N/S (PCIe x8) & 6.6 \\
    us4R \cite{us4usUs4R} & US & 3-lvl 20MHz 180Vpp & 12-bit @80 & up to 32~GB & N/S (2$\times$ PCIe x16) & \textcolor{darkgreen}{\textbf{$\approx$24\textsuperscript{\textdagger}}} \\
    OpenSonics \cite{opensonicsUltrasoundResearchProducts} & US+OA & 3-lvl 15MHz 200Vpp & 14-bit @50 & N/S*\textsuperscript{\textdagger} & Yes (4$\times$ 10Gb LAN) & 4 \\
    CONUS \cite{lagonigroModularScalableSystem2025} & US & \textcolor{darkgreen}{\textbf{5-lvl 50MHz 200Vpp}} & 10-bit @50 & 8~GB*\textsuperscript{\textdagger} & Yes (16$\times$ 10Gb LAN) & 9.5 \\
    Robin et al. \cite{robinDualModeVolumetricOptoacoustic2022} & OA+US & 3-lvl 40Vpp & 12-bit @ 40 & N/S & No & 0.125\textsuperscript{\textdagger} \\
    \textbf{This work} & OA+US & \textcolor{darkgreen}{\textbf{5-lvl 50MHz 200Vpp}} & \textcolor{darkgreen}{\textbf{16-bit @125}} & 8GB*\textsuperscript{\textdagger} & \textcolor{darkgreen}{\textbf{Yes (100GbE RDMA)}} & \textcolor{darkgreen}{\textbf{11.95}} \\
    \bottomrule
  \end{tabular}

  $^1$ All rates in MSPS; ``N/S'' = not specified.\\
  $^2$ Maximum on-system memory; * = not all available for raw RF storage. \noindent\textsuperscript{\textdagger} relies primarily on PC side buffering \\
  $^3$ Sustained pre-beamforming data transfer over high-speed link.\\
  $^4$ Measured or reported actual data streaming rate or \noindent\textsuperscript{\textdagger} theoretical maximum for the interface.
\end{table*}

\subsection{US phantom}

The US functionality was validated using a commercial ultrasound phantom (Model 040GSE, CIRS, USA).
The ultrasound transducer was driven in plane-wave 0-degree transmission mode, with a 3-pulse excitation at a 5 MHz center frequency. Data from the 16 channels was acquired at a sampling frequency of 80 MSPS.
For dataflow validation, the stream was expanded to a 256-channel payload by padding 240 channels with dummy
data.

After the trigger signal is received by the pulser, the FPGA packet generator delays the start of acquisition by 60 cycles, then starts streaming the frame data to the circular buffer, as well as triggering the start of RDMA transmission. 
A total of six packets, each 256 KB in size, were transmitted for the acquired frame.

\begin{figure}[ht]
    \centering
    \includegraphics[width=0.8\columnwidth]{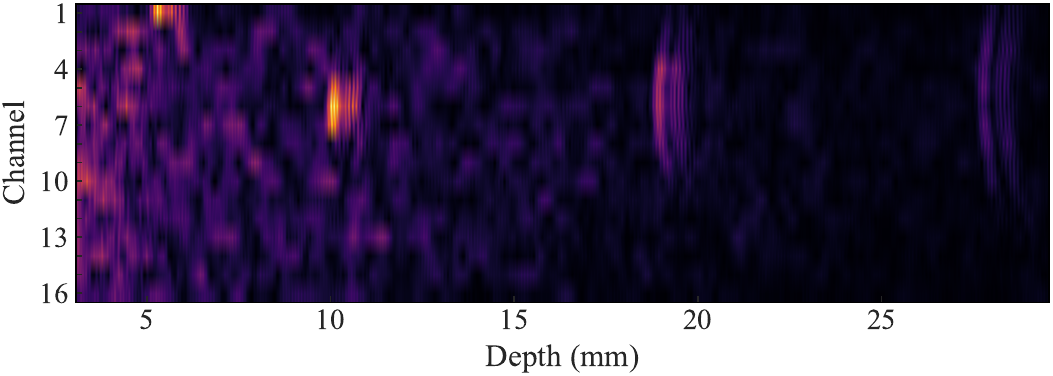}
    \caption{Demonstration phantom US image for the 16-channel validation system, with end-to-end data acquisition and transmission from the FPGA to the workstation via RDMA.}
    \label{fig:end_to_end_acquired_image}
\end{figure}

On the workstation, the received raw data were bandpass filtered, envelope-detected, and displayed as in
Fig.~\ref{fig:end_to_end_acquired_image} to validate the correct reception of the phantom data. 

\subsection{OA phantom}

\begin{figure}[ht]
    \centering
    \includegraphics[width=0.8\columnwidth]{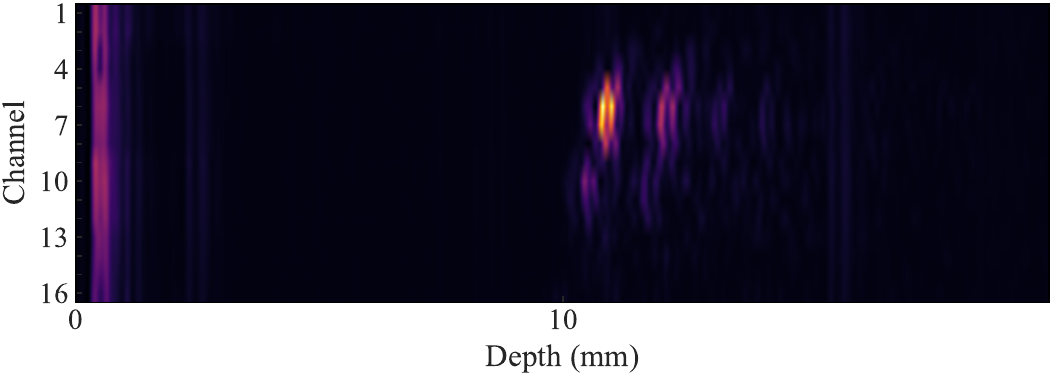}
    \caption{Demonstration phantom OA image for the 16-channel validation system, with end-to-end data acquisition and transmission from the FPGA to the workstation via RDMA.}
    \label{fig:oa_end_to_end_acquired_image_OA}
\end{figure}

To validate the feasibility of LtL for OA imaging, a series of images of a custom-made black tape phantom were acquired using the demonstration system. A function generator was used to simultaneously trigger both the LtL system and a laser diode (LD)-based OA excitation source (QD-Qxy10-ILO, Quantel laser, France). The LD emitted pulsed light at a wavelength of 808 nm, delivering a pulse energy of 1 mJ with a pulse width of 100 ns. The ultrasound transducer and the light source were positioned on the same side of the black tape. A layer of transparent Humimic \#0 gelatine was placed between the transducer and the black tape, serving simultaneously as a light-transmission medium and an ultrasound-coupling layer. Data was acquired as in the US validation, with 16 channels acquired and expanded to 256 channels for transmission over RDMA. The raw OA data received on the host PC are bandpass filtered and envelope detected before image reconstruction. Fig.~\ref{fig:oa_end_to_end_acquired_image_OA} shows a representative OA raw data obtained with the proposed system. The OA signals and their corresponding echoes were predominantly detected by channels 5–8. 

\section{Comparison with State-of-the-Art}
\label{sect:soa}

Table~\ref{tab:platform-comparison} compares LtL to state-of-the-art research platforms with large (256) channel count. LtL stands out as one of the few platforms supporting dual modality (US+OA), together with the OpenSonics \cite{opensonicsUltrasoundResearchProducts} and the work by Robin et al. \cite{robinDualModeVolumetricOptoacoustic2022}. Additionally, to support the wide bandwidth requirements of OA, also towards novel applications demanding high-frequency transducers \cite{haedickeHighresolutionOptoacousticImaging2020}, LtL and the latest Verasonics NXT are the only platforms supporting sampling speeds as high as 125 MSPS.

Regarding data transmission, while multiple platforms rely on a PCIe card, which requires a dedicated workstation for the connection, LtL follows a networking approach by relying on optical-links and data transmission via Ethernet. Notably, by exploiting RDMA communication, LtL achieves SoA sustained peak datarates (as high as 11.95 GB/s), without being limited by the buffer sizes but only by the available memory at the host PC.

Figure \ref{fig:soa_radar} summarizes the comparison in a radar plot, showing that LtL outperforms prior works in most categories, particularly in terms of AFE and pulser performance, and streaming bandwidth. Scaling in the plot was performed based on the following criteria:
\begin{itemize}
    \item \textbf{Modality}: dual OA+US vs just US modality.
    \item \textbf{Pulser}: equal weighting is given to voltage (higher is better) and number of levels (5 is better than 3).
    \item \textbf{AFE/ADC}: equal weighting is given to bit depth (higher is better) and sampling rate (higher is better).
    \item \textbf{Buffer}: maximum on-system memory is considered, with higher values being better. Systems that rely primarily on host-side buffering are assigned a buffer size equal to the maximum on-system memory (ULA-OP 256).
\end{itemize}

\section{Conclusion and Outlook}
\label{sect:conclusion}

This work introduces ListenToLight (LtL), a software-defined architecture for ultrafast OA and US imaging. LtL addresses key limitations of existing OA–US systems by combining wideband, low-noise analog front-ends with a Zynq UltraScale+ MPSoC that tightly integrates FPGA fabric and an APU for fully software-defined acquisition. A 100-GbE RDMA backend enables sustained raw-data streaming at state-of-the-art peak datarates (11.95 GB/s), overcoming the buffering and burst-transfer constraints that limit long-duration ultrafast acquisitions on most existing platforms.

This paper validated the core building blocks of LtL using a 16-channel demonstration system assembled from commercial evaluation boards. The demonstration prototype showcases the complete JESD204B reception chain, framing logic, circular buffering, and high-throughput RDMA transmission to the host PC. Finally, US and OA phantom measurements confirmed the correct system operation under realistic data acquisition scenarios.

\subsection{Scalability to larger channel count}
\label{sect:scalability}

The 16-channel demonstration system was designed to validate the architecture and main building blocks of the ListenToLight platform while guaranteeing scalability to larger channel count.
Most of the digital backend operates at the full 256-channel width, including the JESD204B, framing logic, circular buffers, RDMA streaming, and APU-based control. Further scaling the platform mainly requires:
\begin{itemize}
    \item {Front-end scaling}: replicating sixteen-times the 16-channel front-end (based on the STHV1600 pulser and AFE58JD48 AFE). The interfaces exposed by the FPGA for the control and data acquisition from these devices are sufficient to enable this scale-up. 
    \item {FPGA scaling}: the JESD204B RX PHY and ListenToJESD cores scale linearly with the number of lanes (2 JESD lanes per 16-channel group). The available resources (shown in Tab.~\ref{tab:fpga-utilization}) are sufficient for the scaling to 256 channels.
    \item {Multi-module synchronization}: multiple AFE devices sharing a global SYSREF and AFECLK clock tree, ensuring deterministic latency across modules.
\end{itemize}

Future work will address these points by designing integrated acquisition cards and upgrading the FPGA design for 256-channel operation.

\section*{Acknowledgment}

The authors thank Alfonso Blanco Fontao, Hansjoerg Gisler, and Thomas Quanbrough (ETH Z{\"u}rich) for technical support. The authors acknowledge support from the ETH Research Grant ETH-C-01-21-2 (Project ListenToLight).

\bibliographystyle{IEEEtran}

\end{document}